\documentstyle[aps,twocolumn,floats,epsf]{revtex}
\begin{document}
\draft
\title{Nanosized superconducting constrictions.}
\author{M. Poza$^{\dag }$, E. Bascones$^{\dag ,\ddag }$, J.G. Rodrigo$^{\dag },$ N. Agra{%
\"\i }t$^{\dag }$, S. Vieira$^{\dag }$ and F. Guinea$^{\ddag }$}
\address{$^\dag$ Laboratorio de Bajas Temperaturas,
Instituto Nicol\'as Cabrera,
\\ Departamento de F\'{\i}sica de la
Materia Condensada C-III,\\ Universidad Aut\'onoma de Madrid,
E-28049 Madrid. Spain.\\ $^\ddag${I}nstituto de Ciencia de
Materiales de Madrid,\\ CSIC, Cantoblanco, E-28049 Madrid, Spain.}
\date{\today}
\maketitle

\begin{abstract}
Nanowires of lead between macroscopic electrodes are produced by means of an
STM. Magnetic fields may destroy the superconductivity in the electrodes,
while the wire remains in the superconducting state. The properties of the
resulting microscopic Josephson junctions are investigated.
\end{abstract}


%\twocolumn[\hsize\textwidth\columnwidth\hsize\csname@twocolumnfalse%
%\endcsname

\narrowtext
It is well known that the magnetic field supresses
superconductivity in type I materials only if the dimensions of the
sample are sufficently large, compared to the coherence
length\cite{dG66,Ti96}. Superconductivity survives in small
systems. A number of experiments showed the enhancement of the
critical field in thin films\cite{MD64}. Advances in nanotechnology
have made possible to study this effect in small metallic
particles\cite {Metal94,Betal97,Getal97}. A different realization
can be achieved by applying a magnetic field above the bulk
critical value to microscopic constrictions\cite{ARV93,RAV96}.
Then, it can happen that the constriction itself remains
superconducting, while the electrodes become normal\cite
{PRV96,Po97}. This novel device allows us to probe
superconductivity at small scales through transport measurements.
Properties like the critical current, and its dependence on field
and thickness, can be studied in detail. If the superconducting
properties are not homogeneous along the constriction, it can
behave as a Josephson junction of nanoscopic dimensions.

Narrow constrictions are generated by pressing an STM tip made of a
metal which remains normal at low temperatures (Pt-Ir or Au) into a
Pb substrate. The substrate has an area of approximately 1 cm$^2$,
and a thickness of 0.5 mm. When the tip is first pressed into the
substrate, it gets covered by lead atoms. Upon succesive raising
and lowering of the tip, a lead bridge is formed between the tip
and the substrate. The aspect ratio of this bridge can be varied by
changing the position of the tip in a controlled way and its size
can be estimated from the evolution of the conductance during this
process as detailed in ref.\cite{U97}. The advantage of working
with normal tips is that the magnetic field needs only to be
applied to the substrate and the constriction. A magnetic field of
2.6 kG (approximately five times the zero temperature critical
field of lead) at the surface of the sample was produced by a small
permanent magnet placed under the sample.

%******
\begin{figure}
\vspace{0mm}
\begin{center}
\leavevmode
    \epsfxsize=60mm
 \epsfbox{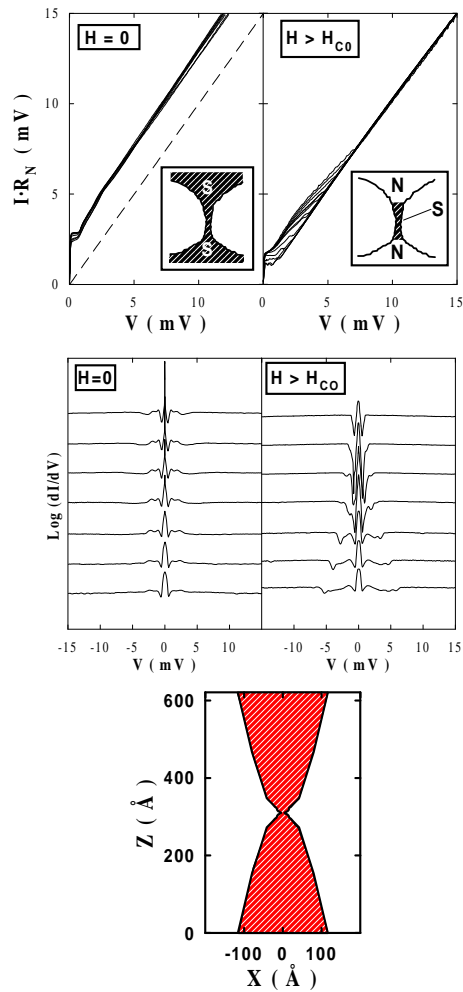}
 \end{center}
\vspace{0in}

\caption[]{Top panel: I-V characteristics of Pb constrictions with and without an
applied field. The inset shows a sketch of the expected situation
at the constriction. The applied field is five times the bulk
critical field of lead. Middle panel: conductances for the same
constrictions as at the top part. Bottom panel: estimated
dimensions of a typical constriction following the procedure of
ref. \protect\cite{U97}.}
\label{IV}
\end{figure}
%********

I-V characteristics for different constrictions are shown in
Fig.\ref{IV}, along with a sketch of the typical dimensions of the
device. The curves show clear signatures of the Josephson effect,
with and without the magnetic field.

The conductance at high voltages is determined by the shape of the
constriction. It is given, approximately, by $2 e^2/h$ times the
number of channels which can be acommodated within the narrowest
part. The number of channels, in turn, goes as the cross section
over an area of atomic dimensions. Hence, the high voltage
conductance gives a measurement of the cross section of the
constriction.

At zero voltage, the Josephson effect shunts the constriction,
leading to the observed peak in the conductivity. The residual
resistivity is due to scattering in the normal parts, outside the
constriction. At finite voltages, but below the gap of the
superconducting region, resonance processes due to Andreev
scattering are observed.

%******
\begin{figure}
\vspace{0mm}
\begin{center}
\leavevmode
    \epsfxsize=70mm
 \epsfbox{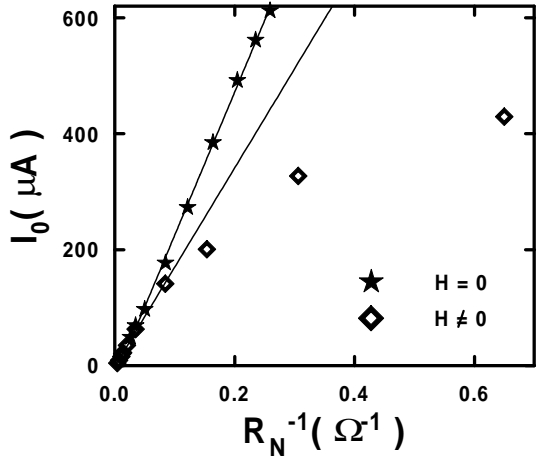}
 \end{center}
\vspace{0in}

\caption[]{Values of the critical current, versus the high voltage
coductance, for different constrictions. The applied magnetic field
is five times the critical magnetic field of Pb.}
\label{ic}
\end{figure}
%********

Figure \ref{ic} shows the critical current versus normal state
conductance with and without a magnetic field. In the absence of an
applied field, both quantities are proportional. This can be
understood because the critical current of a constriction measures
the number of conducting channels within it.

A magnetic field above the bulk critical value reduces the critical
current. The effect is more pronounced in the wide constrictions.
In order to analyze this effect, we calculate the free energy of a
cylindrical superconductor in the presence of a field, which also
supports a current. Let us assume that the field is constant within
the cylinder. This situation describes well cylinders much narrower
than the penetration depth, $r \ll \lambda ( T )$. The current
carried by the condensate is proportional to the gradient of the
phase of the order parameter, which we write as $| \psi | e^{i
\phi}$. Then, the free energy, per unit length is:
\begin{equation}
g = \pi r^2 \left[ \left( \alpha + \frac{\hbar^2}{2 m} | \nabla
\phi |^2
\right) | \psi |^2 + \frac{\beta}{2} | \psi |^4 + \frac{e^2 r^2}{8 m c^2}
H^2 | \psi |^2 \right]  \label{freeenergy}
\end{equation}
where $\alpha , \beta$ and $m$ follow the standard notation\cite{dG66}.

Minimizing with respect to $|\psi^2 |$, we obtain:
\begin{equation}
| \psi |^2 = | \psi_0 |^2 \left( 1 - \frac{\frac{\hbar^2}{2 m} |
\nabla \phi |^2 + \frac{r^2 e^2}{8 m c^2} H^2}{| \alpha |} \right)
\label{op}
\end{equation}
where $\psi_0^2 = \frac{\alpha}{\beta}$. In addition, $\alpha = \frac{2 e^2}{%
m c^2} H_c^2 ( T ) \lambda^2 ( T )$. Hence,
\begin{equation}
| \psi |^2 = | \psi_0 |^2 \left( 1 - \frac{\hbar^2 | \nabla \phi |^2}{2 m |
\alpha |} - \frac{\epsilon^2 h^2}{16} \right)  \label{op1}
\end{equation}
where $\epsilon = \frac{r}{\lambda (T)}$ and $h = \frac{H}{H_c
(T)}$. Finally, we relate $\nabla \phi$ to the current flowing
along the cylinder. The critical current, in the absence of a
field, of a cylinder of radius $r$ is\cite{Ti96} $I_c ( T ) =
\frac{ H_c ( T ) r^2 c} {3 \sqrt{6} \lambda ( T )} $. In terms of
$f = \frac{\psi}{\psi_0}$, we can write:
\begin{equation}
f^2 = 1 - \frac{h^2 \epsilon^2}{16} - \frac{4}{27} \frac{i^2}{f^4}
\label{op2}
\end{equation}
where $i = \frac{I}{I_c}$.

Equation (\ref{op2}) allows us to determine the critical field of a
cylinder with no current flowing: $H_{cyl} ( T ) = 4 H_c ( T )
\frac{\lambda ( T )}{r} $. This formula is valid for $r < \lambda (
T )$. For sufficiently large values of $r$, solutions with vortices
threading the cylinder are also possible\cite{MQ97,BZ97}.

In the presence of a current, we obtain the critical field by first
extracting $h ( f , i )$ from (\ref{op2}) and then calculating the value of $%
f$ which maximizes $h$. In this way, we find: $H_{cyl} ( I , T ) = H_{cyl} (
T ) \sqrt{1 - \left( \frac{I}{I_c ( T )} \right)^{2/3}}$. Analogously,
\begin{equation}
I_c ( H , T ) = I_c ( T ) \left[ 1 - \left( \frac{H}{H_{cyl} ( T )}
\right)^2 \right]^{3/2}  \label{icr}
\end{equation}
The transition to the normal state is discontinuous, as in a thin film\cite
{dG66}. When the current reaches its critical value, the superconducting
order parameter in the constriction jumps to zero.

As a function of the radius of the constriction, eq. (\ref{icr})
predicts that the critical current shows a maximum at
$\frac{r^2}{\lambda^2 (T)} =
\frac{32 H}{5 H_c (T)}$. For $H \approx 5 H_c$, we have $r \ge \lambda$. The
present calculations need to be modified, because the field will
not be homogeneous within the constriction. Note, however, that the
maximum field compatible with superconductivity is not much larger
than $\sim H_c$ for a constriction with $r \sim \lambda$. Hence,
the maximum in $I_c ( r )$ cannot be reached, if the applied field
is much larger than $H_c$. It is interesting to note that, in a
cylinder of variable width, regions threaded by vortices and
regions with no vortices can coexist. As the symmetry of the order
parameter is different in each region, a phase boundary should be
generated, and the critical current will be suppressed.

The general trend of $I_c$ as function of $r$ is consistent with
the results shown in fig.(\ref{ic}). In the presence of a field, we
find that $I_c
\propto r^2$ for narrow constrictions, $r \ll \lambda ( T )$. For wider
constrictions, $I_c$ is strongly supressed by a field.

%******
\begin{figure}
\vspace{0mm}
\begin{center}
\leavevmode
    \epsfxsize=70mm
 \epsfbox{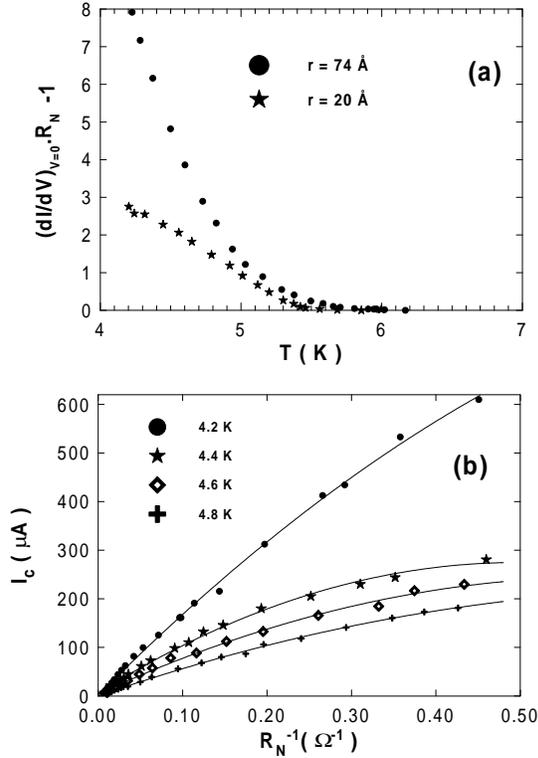}
 \end{center}
\vspace{0in}

\caption[]{(a) Plot of $(dI/dV)_{V=0} R_{N}-1$ vs $T$ for two
different constrictions with estimated radius of 74 \AA\ (circles)
and 20 \AA\  (stars). (b) Critical current vs constriction high
voltage conductance at different temperatures.}
\label{temp}
\end{figure}

%********

Finally, we have studied the dependence of $I_c$ on temperature.
The theory presented here predicts that the critical temperature,
in the presence of an applied field above the bulk critical value,
should depend on the radius of the constriction. In fig.
\ref{temp}a we have represented, for two different constrictions,
the temperature dependence of $(dI/dV)_{V=0} R_{n}-1$, which must
go to zero as we approach the critical temperature. We are aware
that in some circumstances the effect of thermal fluctuations on
the properties of the superconducting region can limit the
significance of the critical temperatutre deduced from fig.
\ref{temp}a. However, it is clear from fig. \ref{temp}b that the
value of the critical current shows a marked dependence on
temperature consistent with the existence of superconductivity in
the region
of the constriction. The full dependence of $%
I_c$ on $T$ and $r$ is:
\begin{eqnarray}
I_c &= &\frac{H_c (0) r^2 c}{3 \sqrt{6} \lambda (0)} \sqrt{\frac{1 - \left(%
\frac{T}{T_c}\right)^2}{1 + \left(\frac{T}{T_c}\right)^2}}  \nonumber \\
&\times &\left[ 1 - \frac{H^2 r^2}{16 H_c^2 (0) \lambda^2 (0)}
\frac{1 +
\left(\frac{T}{T_c}\right)^2}{1 - \left(\frac{T}{T_c}\right)^2} \right]^{%
\frac{3}{2}}  \label{ict}
\end{eqnarray}
This expression predicts that $I_c$ has a linear dependence on
$r^2$ at low temperatures, and curves downward as the temperature
is increased, in agreement with fig. \ref{temp}b.

In conclusion, we present new devices in which to study superconductivity at
small scales. In contrast to previous work, our superconducting regions are
strongly coupled to its environment, which greatly facilitates the
measurement of transport properties. The results presented here are in
agreement with the expected behavior for a narrow superconducting cylinder
in an applied field, in particular, the dependence of critical currents on
width.

This work was done with support from CICyT (Spain) through grant PB96-0875.

\end{document}